\documentstyle[art12]{article}

\begin{document}

\author{R.Yu.~Abdulsabirov${}^1$, A.A.~Bukharaev${}^2$, M.R.~Zhdanov${}^1$,\\
R.Sh.~Zhdanov${}^1$,A.V.~Klochkov${}^1$,S.L.~Korableva${}^1$,\\V.V.~Naletov$%
{}^1$, N.I.~Nurgazizov${}^2$, M.S.~Tagirov${}^1$, D.A.~Tayurskii${}^1$  \\ 
%EndAName
${}^1${\it Kazan State University, Kazan, 420008, Russia}\\${}^2${\it Kazan
Physical-Technical Institute, 420029, Kazan, Russia}}
\title{Detection of nanocracks on double fluoride rare earth crystal surface }
\date{PACS{\it : 61.16.Ch, 67.55.-s, 68.35.Bs, 76.60.-k}}
\maketitle

\begin{abstract}
Predicted earlier, microcracks on the crystal surface of both finely
dispersed $LiYF_4$ powders and single crystals of the Van Vleck paramagnet $%
LiTmF_4$ were detected by using the NMR Cryoporometry and Atomic-Force
Microscopy technique.
\end{abstract}

In paper \cite{1}, in order to explain the specific features of the liquid $%
^3He$ magnetic relaxation in its contact with single crystals of the
dielectric Van Vleck paramagnet $LiTmF_4$ and its diamagnetic analog - $%
LiYF_4$, we suggested the following hypothesis: On the surface of these
crystals, microcracks exist with an average size near 10 nm. In these
microcracks the character of diffusion motion of $^3He$ atoms was supposed
to change drastically. In such a bounded geometry, the atoms of $^3He$
determine, in fact, the kinetics of magnetic relaxation of the liquid $^3He$
which is in contact with a solid body substrate. In this work we present our
results on the detection of the microcracks, their size being near 30nm
(called ''nanocracks''). This was obtained as the result of the two
independent methods - the NMR Cryoporometry and the Atomic-Force Microscopy
(AFM).

For investigation by means of the NMR Cryoporometry, the three finely
dispersed LiYF4 samples with typical sizes: 700 $nm$ (sample-I), 900 $nm$
(sample-II), and 4000 $nm$ (sample-III) were prepared. The technique of
powder preparation can be found in \cite{2}. All samples were immersed in
their own containers, the packing factor being about 0.5. The prepared
samples can be considered as a porous body containing a three-dimensional
net of channels of various dimensions and forms, i.e., in fact, it can be
considered as a single pore with a very complicated geometry. For analysis
purposes such a space is usually partitioned into a set of pores of various
sizes, the pores being assumed to be connect to each other. In this
situation, by the term ''pore size'' the least distance between opposite
walls of a pore is understood. The distribution of the pores with respect to
their sizes is usually determined by means of the gas absorption and mercury
porometry methods. However, both the methods require long time measurements,
rather complicated procedures, and precise experimental equipment. On the
other hand, it is well known that the physical properties of a liquid
bounded by microscopic pores strongly differ from those of a bulk liquid
(see, for example, \cite{3},\cite{4}). In part, the melting temperature can
lower due to increasing contribution from the relative part of the surface
free energy ( for example, the melting temperature of $H_2O$ may occur 60 $K$
lower in a media with average pores about 2 $nm$, see \cite{5}). This
phenomena lies in the base of pores distribution measuring by means of the
NMR of water protons. The part of the liquid phase and the corresponding
size of pores are determined by the intensity of the pulse-NMR of protons
with relatively greater spin-spin relaxation time $T_2$. Obtained in this
manner, the dependence of the intensity of the NMR line on the temperature
then can be transformed into a curve of pores distribution with respect to
their size with help of the probe curve ''change of melting temperature-
pore size'' (see \cite{6}).

If we start with the assumption that powder particles are solid spheres with
diameter 1000 $nm$, then, for a hexagonal dense packing, our estimates show
that the pores whose size is less than 10 $nm$ constitute a part which is
less than 1\% of the whole quantity. Therefore, we can assert{\it \ a priori}
that greater values of the specific weight will demonstrate the presence of
developed microrelief on the surface of crystal particles.

The curves of the pore size distribution were obtained by using the data of
the impulse NMR of the distilled water protons on the frequencies 20 $MHz$
(a home-made spectrometer) and 80 $MHz$ (Bruker NMR spectrometer W80) for
all three samples. This data is shown on the Fig. 1a. Obviously, if the
nanocracks have no place, the graph distribution has to have a functional
dependence of the type $D^\alpha $, where $D$ stands for the characteristic
size of pores, and the exponent $\alpha $ exceeds 1. Our experimental data
show that the nanocracks are evidently present on the surface of particles.
As one can see on Fig 1b, the distribution has two maximums which correspond
to the specific porous sizes of 3-5nm and 15-20nm, their part occupying near
20\% of the whole hollow volume of the sample.

To visualize the microrelief of a particle surface, we carried out
investigations of the particles of the sample-III by means of the
Atomic-Force Microscope P4-SPM-MDT (produced by NT-MDT, Zelenograd, RF). The
construction of the microscope allows to assert the presence of the
nanocracks with a size greater than 30 $nm$. The AFM-images of various
magnifications of a particle surface can be seen on Fig. 2. It is evidently
seen that the surface possesses sufficiently developed microrelief.

One may suppose that the nanocracks appear due to the technique of powder
preparation, namely, in view of the mechanical grinding. In this connection
the investigation of the surface of a single crystal sample seems to be of
interest. In view of our further investigations of the magnetic coupling
between the liquid $^3He$ and a solid-state substrate, we applied the
AFM-investigation to a single crystal of $LiTmF_4$ . The surfaces (100),
(110), and (001) of the single crystal were mechanically polished in a
certain direction with respect to crystallographic axes (the abrasive for
coarse polishing had a grain 1030, while the finishing was made with the
GOI-paste with particles smaller than 0.5). One of the AFM-pictures of the
single crystal surface is shown on the Fig 3. Our experimental results do
not show a correlation between the nanocracks' character and
crystallographic planes. Obviously, the obtained surface microrelief cannot
be conditioned as the only result of a mechanical polishing in a certain
direction. Thus, investigations detect the presence of the nanocracks on the
surface of both particles and single crystals. The most probable reason for
the nanocracks' appearance is formation of local mechanical stresses in a
sample sawing by a diamond-charged saw, which then are discharged due to
surface reforming.

We should note that the crystal structure of double fluorides of rare earth
has no cleavage planes. The AFM experiments carried out with single crystal $%
LiF$ and $CaF_2$ samples which have evident cleavage planes (100) and (111),
respectively, showed the absence of nanocracks on the surfaces of a
''fresh'' cleavage.

The investigations carried out result in the established presence of
nanocracks, predicted earlier by the authors (see \cite{1}), and their
distribution by size on $LiYF_4$ and $LiTmF_4$ crystal surface. This seems
to be of importance for both the detailed investigation of the nature of
magnetic coupling between liquid $^3He$ and a solid-state substrate, on one
hand, and the practical realization of a method of dynamic nuclei
polarization of liquid $^3He$ with the use of the dielectric Van Vleck
paramagnets. This technique was suggested by the authors in \cite{7}. From
our results it follows that the surface of particles of finely dispersed
powders is deformed in a high degree. The latter is in good correspondence
with our estimates of the deformed layer part in systems of this kind (see 
\cite{8}).

We wish to express gratitude to Prof.\frame{M.A.Teplov} for constant
interest to our investigations, to Prof. T.Ando (Kanazawa University, Japan)
for his preliminary AFM NMR investigations and to P.P Chernov for his
support in the NMR investigations.

The work was supported by Russian Foundation for Basic Research (the
projects 96-02-16323 and 97-02-16470).

\newpage\

{\bf Figure capture}

Fig.1. a) The relative volume of nanocracks in dependence on their size for
three samples, obtained by NMR Cryoporometry. b) The distribution of
nanocracks with respect to their size.

Fig.2. a), b) The picture of the surface of a particle of sample III,
obtained by AFM; one division of the graduation scale is 100 $nm$.

Fig.3. The picture of the surface (110) of a single crystal $LiTmF_4$,
obtained by AFM. Polishing was done along the $X$- axis, scanned along the $%
X $-axis. The graduation on both the axes $X$ and $Y$ is 1000 $nm$, on the
axis $Z$ -100 $nm$.

\end{document}